\documentclass[a4paper]{jpconf}
\usepackage{graphicx}
\begin{document}
\title{Recent Advances in the Application of the Shell Model Monte Carlo Approach to Nuclei}

\author{Y. Alhassid$^{1}$,  M. Bonett-Matiz$^{1}$, A. Mukherjee$^{2}$, H. Nakada$^{3}$ and C. \"Ozen$^{4}$}

\address{$^{1}$Center for Theoretical Physics, Sloane Physics Laboratory, Yale University, New Haven, Connecticut 06520, USA\\ 
$^{2}$ECT*, Villa Tambosi, I-38123 Villazzano, Trento, Italy\\
$^{3}$Department of Physics, Graduate School of Science, Chiba University, Inage, Chiba 263-8522, Japan\\
$^{4}$Faculty of Engineering and Natural Sciences, Kadir Has University, Istanbul 34083, Turkey}

\ead{yoram.alhassid@yale.edu}

\begin{abstract}
The shell model Monte Carlo (SMMC) method is a powerful technique for calculating the statistical and collective properties of nuclei in the presence of correlations in model spaces that are many orders of magnitude larger than those that can be treated by conventional diagonalization methods. We review recent advances in the development and application of SMMC to mid-mass and heavy nuclei. 
\end{abstract}

\section{Introduction}

Most microscopic studies of heavier nuclei are based on mean-field methods such as density functional theory.  However, important correlation effects beyond the mean field can be missed.  The configuration-interaction (CI) shell model approach accounts for correlations and shell effects but conventional diagonalization methods are limited by the size of the many-particle model space.  The auxiliary-field Monte Carlo (AFMC) method  enables calculation of thermal observables and ground state properties in model spaces that are many orders of magnitude larger than model spaces that can be treated by conventional diagonalization methods.  In nuclear physics this method  is also known as the shell model Monte Carlo (SMMC) method~\cite{Lang1993,Alhassid1994}. SMMC has proven to be a powerful method in calculating statistical and collective properties of nuclei~\cite{Nakada1997,Alhassid1999,Alhassid2007,Alhassid2008,Ozen2013}. 

Here we review recent advances in the development and application of SMMC to nuclei.  In particular, we describe a  method to circumvent the sign problem that arises from the projection on an odd number of particles and its application to calculate pairing gaps in mid-mass nuclei and level densities of odd-mass isotopes. We also discuss the application of SMMC to describe the microscopic emergence of collectivity in heavy rare-earth nuclei and present results for the collective enhancement factors of level densities in such nuclei.

\section{Auxiliary-field Monte Carlo (AFMC) method}\label{AFMC}

The AFMC method utilizes the Hubbard-Stratonovich transformation~\cite{Hubbard1957}, in which the imaginary-time propagator $e^{-\beta H}$ of a nucleus that is described by a Hamiltonian $H$ at inverse temperature $\beta=1/T$  is represented by a functional integral of one-body propagators of non-interacting nucleons over external auxiliary fields
\begin{equation}\label{HS}
e^{-\beta H} = \int {\cal D}[\sigma]
G_\sigma U_\sigma \;.
\end{equation}
 Here $G_\sigma$  is a Gaussian weight and $U_\sigma$ is the propagator of non-interacting nucleons moving in time-dependent auxiliary fields $\sigma(\tau)$.   The thermal expectation value of an observable $O$ is 
 \begin{equation}
 \langle O \rangle_T ={{\rm Tr}(e^{-\beta H}O) \over {\rm Tr} e^{-\beta H}} = {\int D[\sigma] W_\sigma \Phi_\sigma \langle O \rangle_\sigma
\over \int D[\sigma] W_\sigma \Phi_\sigma} \;,
 \end{equation}
where $W_\sigma = G_\sigma |\Tr\, U_\sigma|$ is a positive-definite weight function and $\Phi_\sigma = \Tr\, U_\sigma/|\Tr\, U_\sigma|$ is the Monte Carlo sign function. The quantity $\langle O \rangle_\sigma =
 {\rm Tr} \,( O U_\sigma)/ {\rm Tr}\,U_\sigma$ is the thermal expectation value of the observable for a given configuration $\sigma$ of the auxiliary fields. In SMMC, we choose  samples $\sigma_k$ distributed according to $W_\sigma$ and  estimate the observables from
$\langle O\rangle_T \approx  {\sum_k  \langle  O \rangle_{\sigma_k} \Phi_{\sigma_k} / \sum_k \Phi_{\sigma_k}}$.
AFMC often suffers from a sign problem at low temperatures. However, there are classes of good-sign interactions that are free of the sign problem and are composed of the dominant collective components of effective nuclear interactions~\cite{Zuker1996}. Small bad-sign components can be treated by the extrapolation method of Ref.~\cite{Alhassid1994}.  

\section{Circumventing the odd-particle sign problem}\label{odd-particle}

Applications of SMMC to odd-even and odd-odd nuclei have been hampered by a sign problem that originates from the projection on a odd number of particles. Such a sign problem occurs at low temperatures even for good-sign interactions and makes it difficult to estimate accurately the ground-state energy of an odd-particle system. 
A breakthrough was a method we introduced to  calculate accurately the ground-state energy of an odd-particle system that circumvents this sign problem~\cite{Mukherjee2012}.  It is based on the scalar imaginary-time single-particle Green's functions 
\begin{equation}
G_\nu(\tau) = \sum_m \langle \mathcal{T} a_{\nu m}(\tau) a^\dagger_{\nu m} (0)\rangle \;.
\end{equation}
Here $\mathcal{T}$ denotes time ordering, $\nu = (n l j)$ denotes the radial quantum number $n$, orbital angular momentum $l$ and total spin $j$  of the single-particle levels,  and $m$ is the magnetic quantum number of $j$.  In an asymptotic regime of sufficiently large $\tau$ but $\tau \ll \beta$, the Green's functions of an even-even nucleus $A$ have the form
$G_\nu(\tau) \sim e^{-\beta[E_j(A\pm 1) - E_{\rm g.s.}(A) ]|\tau|}$, 
where $E_j(A\pm 1)$ is the lowest energy of spin $J=j$ in the neighboring even-odd nuclei $A\pm1$ and $E_{\rm g.s.}$ is the ground-state energy of the nucleus $A$.  Thus $E_j(A\pm 1)$ can be determined from the slope of the calculated $\ln G_\nu$ and the ground-state energy of the even-even nucleus.  The ground-state energy of the even-odd nucleus is then found by minimizing $E_j(A\pm1)$ over $j$. 
\begin{figure}[h]
\center{\includegraphics[width= 0.9\textwidth]{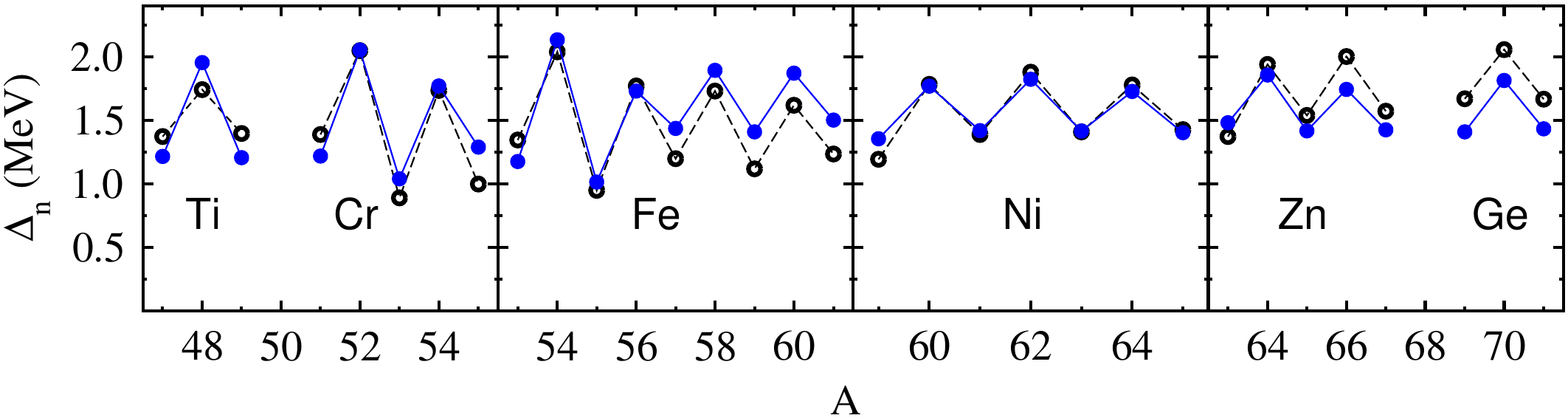}}
\caption{\label{pairing-gaps} Neutron pairing gaps $\Delta_n$ as a function of mass number $A$ in families of $fpg_{9/2}$-shell isotopes. The gaps calculated in SMMC (solid circles) are compared with the experimental gaps (open circles).
Adapted from Ref.~\cite{Mukherjee2012}.}
\end{figure}

We used this method to calculate neutron pairing gaps in mid-mass nuclei from the second-order energy difference in neutron number $\Delta_n = (-)^N [E(Z,N+1) + E(Z,N-1) -2E(Z,N)]/2$, where $E(Z,N)$ is the ground-state energy of the nucleus with $Z$ protons and $N$ neutrons.  Fig.~\ref{pairing-gaps} shows neutron pairing gaps for mid-mass isotopes. The calculations were carried out in the complete $fpg_{9/2}$ shell with the interaction of Ref.~\cite{Nakada1997}. The SMMC results (solid circles) are in overall good agreement with the experimental values (open circles).

\section{Level densities}\label{level-densities}

SMMC has been successful in calculating nuclear state densities $\rho(E_x)$ versus excitation energy $E_x$ in the presence of correlations. The density calculated in SMMC is the {\em state} density, in which each level with spin $J$ is counted $2J+1$ times. However, often the density measured in the experiments is the {\em level} density, in which each level with spin $J$ is counted once.  

We recently introduced a method to calculate the level density directly in SMMC~\cite{Alhassid2013}. Denoting by $\rho_M$ the level density at given value $M$ of the angular momentum component $J_z$, the level density $\tilde \rho$ is given by $\tilde{\rho} = \rho_{M=0}$ for even-mass nuclei and by $\tilde{\rho} = \rho_{M=1/2}$ for odd-mass nuclei.
 $J_z$ projection can be carried out exactly in SMMC as discussed in Ref.~\cite{Alhassid2007}. 

\begin{figure}[h]
\center{\includegraphics[width= 0.7\textwidth]{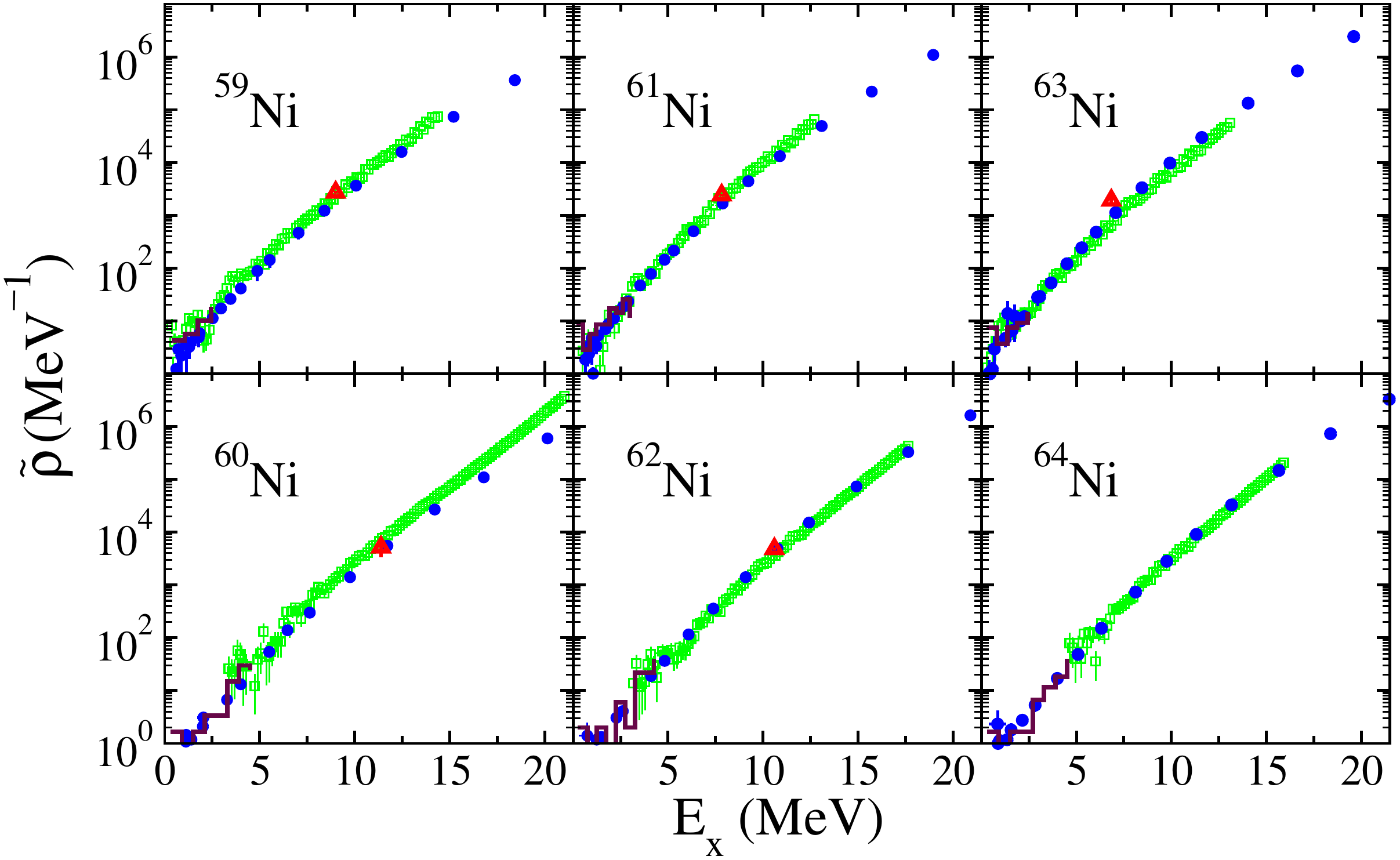}}
\caption{\label{Ni-level} 
Level densities $\tilde \rho$ in a family of nickel isotopes. The SMMC level densities (solid circles) are compared with level densities determined from various experimental data sets: proton evaporation spectra (open squares and quasi-continuous lines), level counting data (histograms) and neutron resonance data when available (triangles). 
Taken from Ref.~\cite{Bonett2013}.}
\end{figure}

Level densities were recently measured by proton evaporation spectra in a family of nickel isotopes  $^{59-64}$Ni~\cite{Voinov2012}. We calculated these level densities~\cite{Bonett2013} within the $fpg_{9/2}$ shell using the Hamiltonian of Ref.~\cite{Nakada1997}. The calculation of the ground-state energies of the odd-mass nickel isotopes, and hence their excitation energies, was made possible by the method discussed in Sec.~\ref{odd-particle}.
Fig.~\ref{Ni-level} shows the level densities of $^{59-64}$Ni. The SMMC results are in close agreement with the level densities determined from various experimental data sets. 

\section{Microscopic emergence of collectivity in heavy nuclei}\label{collectivity}

Heavy nuclei are known to exhibit various types of collectivity.  Typically, nuclei near shell closure are spherical and display vibrational collectivity, while mid-shell nuclei are strongly deformed and display rotational collectivity. Different types of collectivity are well described by empirical models but much less is understood microscopically. 
Here we use SMMC to describe the emergence of collectivity in heavy nuclei within the framework of the CI shell model approach. 

\subsection{Thermal signatures of collectivity}

A given type of collectivity is usually identified by its associated spectrum.  While SMMC is a powerful method for calculating thermal and ground-state properties in very large model spaces, it does not provide detailed spectroscopic information.  To overcome this difficulty, we have identified a thermal observable whose low-temperature behavior is sensitive to the type of collectivity.   Such an observable is  $\langle {\mathbf J}^2 \rangle_T$, where ${\mathbf J}$ is the total angular momentum.  Assuming an even-even nucleus  with a rotational or a vibrational ground-state band and an excitation energy $E_{2^+}$ of the first $2^+$ state, we have~\cite{Alhassid2008,Ozen2013}
\begin{eqnarray}\label{J2-theory}
\langle \mathbf{J}^2 \rangle_T \approx
 \left\{ \begin{array}{cc}
 30  \frac{e^{-E_{2^+}/T}}{\left(1-e^{- E_{2^+}/T}\right)^2} &{\rm vibrational\; band}  \\
 \frac{6}{E_{2^+}} T & {\rm rotational \;band}
 \end{array} \right. \;.
\end{eqnarray}
\begin{figure}[h]
 \center{\includegraphics[width=0.6\textwidth]{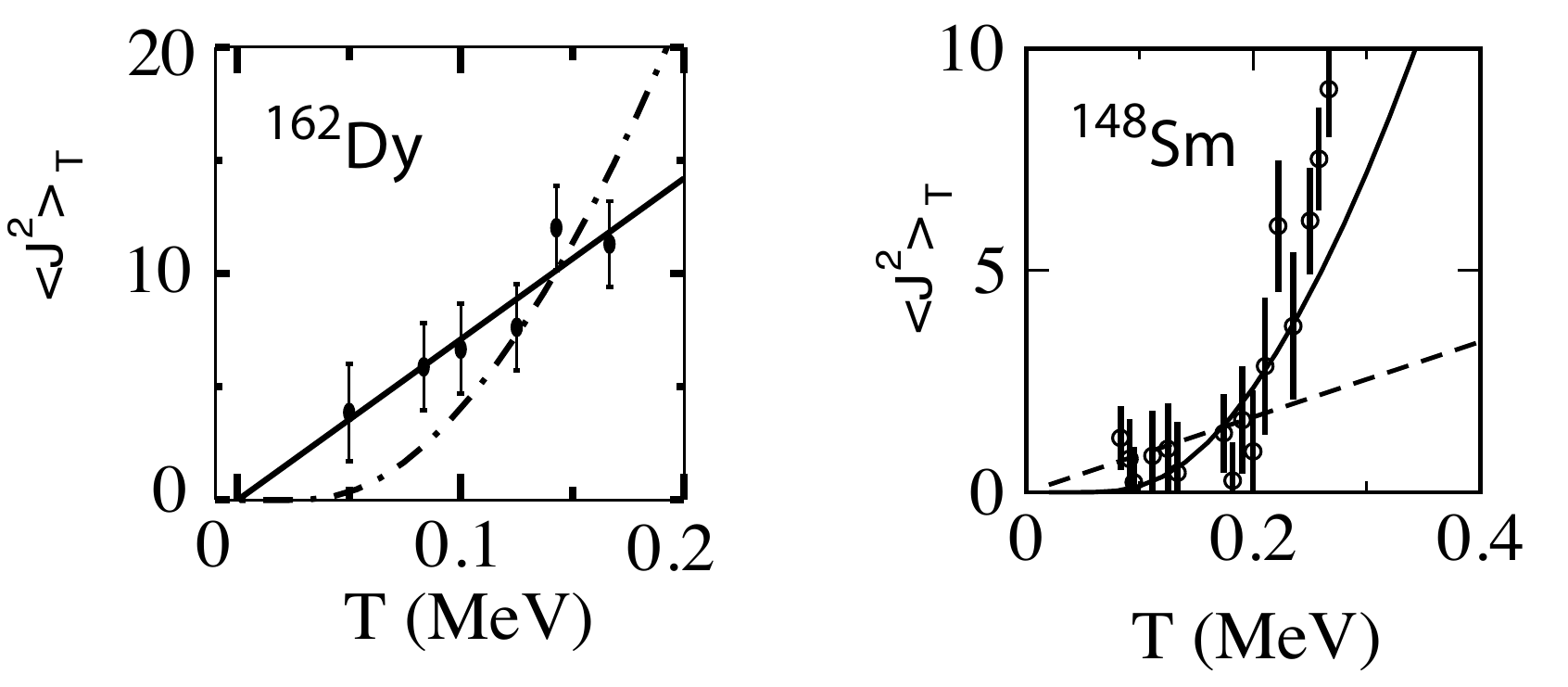}}
 \caption{\label{J2} Left: $\langle \mathbf{J}^2 \rangle_T$  in $^{162}$Dy. The SMMC results (solid circles) are compared with fits to the rotational model (solid line) and vibrational model (dashed-dotted line).  Adapted from Ref.~\cite{Alhassid2008}.    Right: $\langle \mathbf{J}^2 \rangle_T$  in $^{148}$Sm. The SMMC results (open circles) are compared with fits to the vibrational model (solid line) and rotational model (dashed line).}
\end{figure}

We carried out calculations for rare-earth nuclei using the $50-82$ shell plus $1f_{7/2}$ for protons, and the $82-126$ shell plus $0h_{11/2}, 1g_{9/2}$ for neutrons. The Hamiltonians we used are given in Refs.~\cite{Alhassid2008,Ozen2013}. The left panel of Fig.~\ref{J2} shows  $\langle {\mathbf J}^2 \rangle_T$ for the deformed rare-earth nucleus $^{162}$Dy. The SMMC results are compared with fits to both the rotational and vibrational models.  The agreement of the SMMC results with the rotational model confirms that we are able to describe the rotational character of  $^{162}$Dy in the framework of a truncated spherical shell model approach.  The fit to the rotational formula gives $E_{2^+}=84.5\pm 8.9$ keV, in agreement with the experimental result of $E_{2^+}=80.6$ keV.  The right panel of Fig.~\ref{J2} shows $\langle {\mathbf J}^2 \rangle_T$ for the spherical rare-earth nucleus $^{148}$Sm.  Here the SMMC results follow closely the vibrational formula fit with $E_{2^+}=538\pm 31$ keV, in agreement with the experimental value of $E_{2^+}=550$ keV.

\subsection{The crossover from vibrational to rotational collectivity}

Heavy nuclei are known to exhibit a crossover from vibrational to rotational character as the number of neutrons increases from shell closure towards mid-shell. In a mean-field approximation, this is described by a phase transition from spherical to deformed nuclei. We have studied two such families of rare-earth isotopes, $^{148-154}$Sm and  and $^{144-152}$Nd~\cite{Ozen2013}.   

 \begin{figure}[h]
   \center{\includegraphics[clip,angle=0,width= 0.9\columnwidth]{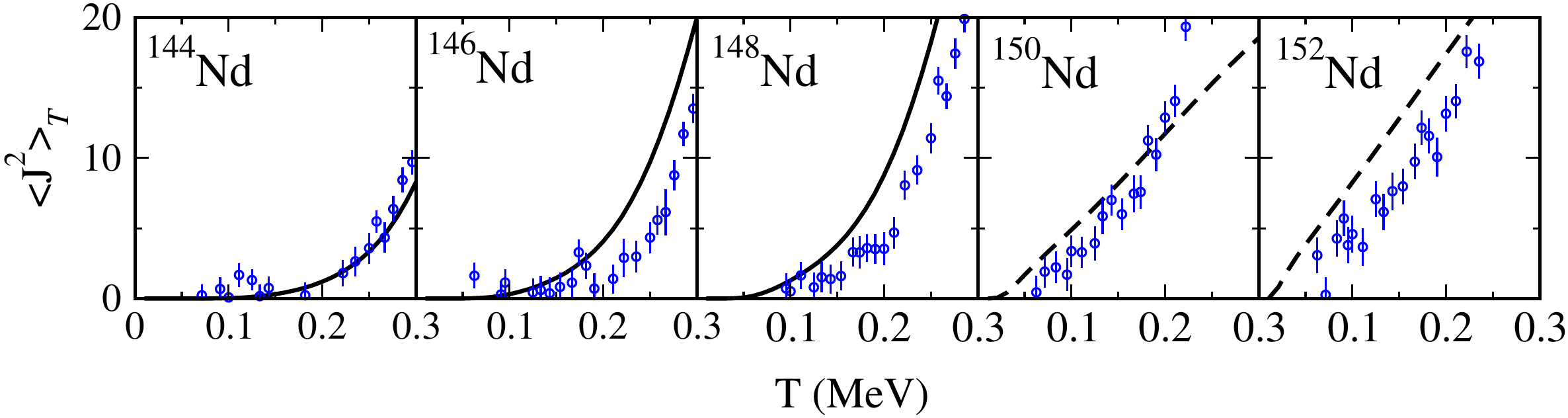}}
    \caption{$\langle \mathbf{J}^2 \rangle_T$ as a function of temperature $T$ for a family of even neodymium isotopes.  The solid lines describe the experimental values extracted using Eq.~(\ref{J2-exp}).  For $^{150-152}$Nd we only use the discrete levels in Eq.~(\ref{J2-exp}) (dashed lines).  Adapted from Ref.~\cite{Ozen2013}. }
  \label{Nd-J2}
\end{figure}

Fig.~\ref{Nd-J2} shows  $\langle {\mathbf J}^2 \rangle_T$ for a family of even neodymium isotopes. The open circles are the SMMC results. We observe a crossover from a soft response to temperature in $^{144-146}$Nd, typical of a vibrational nucleus, to a rigid response in $^{152}$Nd, typical of a rotational nucleus.  The solid lines are extracted from experimental results using 
\begin{eqnarray}
\label{J2-exp}
 \langle \mathbf{J}^2 \rangle_T & = & \frac{1}{Z(T)} \left(\sum_i^N J_i(J_i+1)(2J_i+1)e^{-E_{i}/T} + 
 \int_{E_{N}}^\infty d E_x \: \rho(E_x) \: \langle \mathbf{J}^2 \rangle_{E_x} \; e^{-E_x/T} \right)\;,
\end{eqnarray}
where  $Z(T)=\sum_{i}^{N} (2J_i+1) e^{-E_i/T} + \int_{E_{N}}^\infty d E_x \rho(E_x) e^{-E_x/T}$
 is the corresponding experimental partition function. Here $E_i$ are the measured low-lying energy levels  with spin $J_i$, which we assume to be complete up to to certain threshold energy $E_N$.  Above the threshold energy, we use a backshifted Bethe formula for the level density $\rho(E_x)$ whose parameters are determined from a fit to level counting data at low energy and neutron resonance data at the neutron resonance threshold.
  We use in Eq.~(\ref{J2-exp}) the spin cutoff model estimate $\langle \mathbf{J}^2 \rangle_{E_x} =3\sigma^2$, where  $\sigma^2=I T/\hbar^2$ is calculated assuming a rigid-body moment of inertia $I$.
 We observe an overall good agreement between the SMMC values of $\langle {\mathbf J}^2 \rangle_T$ and the experimentally determined values.  

\section{Collective enhancement factors of level densities in heavy nuclei}

SMMC state densities in rare-earth nuclei were found to be in good agreement with experimental data~\cite{Alhassid2014}. Collective states in these nuclei lead to enhancement of their state density,  which is described by a collective enhancement factor $K$. The decay of $K$ with excitation energy is one of the least understood topics in the modeling of level densities~\cite{Capote2009}, and it is often parameterized by empirical formulas.  We define a collective enhancement factor by
$K= {\rho_{\rm SMMC} / \rho_{\rm HFB}}$, 
where $\rho_{\rm SMMC}$ is the total SMMC state density and  $\rho_{\rm HFB}$ is the level density calculated in the finite-temperature Hartree-Fock-Bogoliubov (HFB) approximation.  The latter describes the density of intrinsic states and thus the above ratio is a measure of the enhancement in the number of states due to collective states that are built on top of the intrinsic states.
\begin{figure}[t]
   \center{\includegraphics[clip,angle=0,width= 0.9\columnwidth]{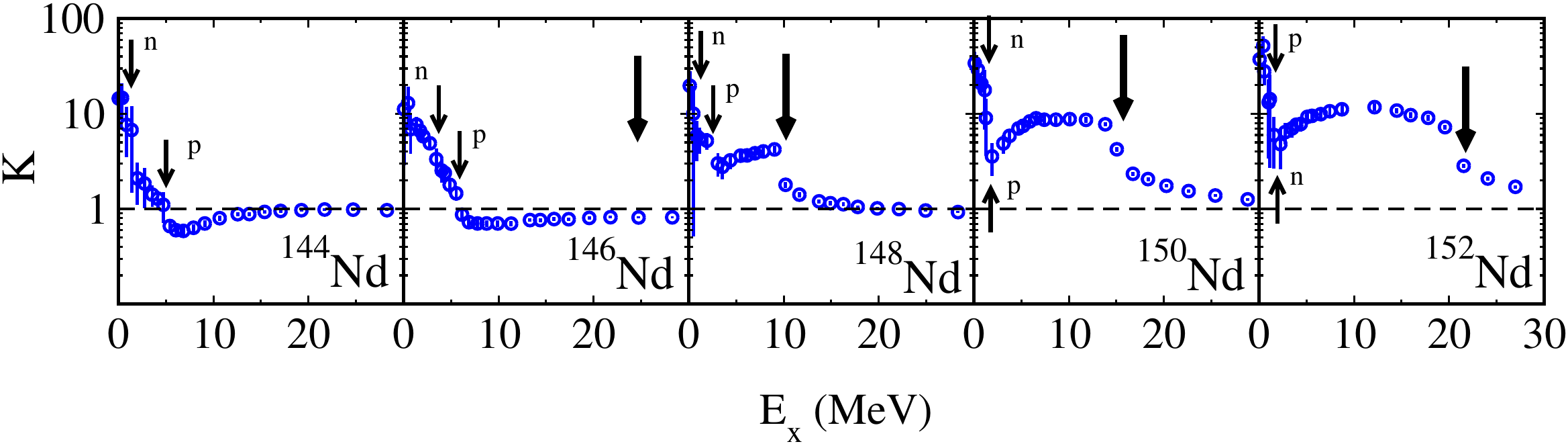}}
    \caption{Collective enhancement factor $K$ (open circles) of the even neodymium isotopes  $^{144-152}$Nd versus excitation energy $E_x$ (see text for the definition of $K$).  The thin arrows correspond to the proton and neutron pairing transitions, while the thick arrows describe the shape transitions.  Adapted from Ref.~\cite{Alhassid2014}.}
    \label{Nd-enh}
\end{figure}

Fig.~\ref{Nd-enh} shows the calculated enhancement factor $K$ as a function of excitation energy $E_x$ for the even neodymium isotopes $^{144-152}$Nd. We also show by arrows the energies of the various thermal phase transitions that occur in the HFB approximation. The thin arrows correspond to the pairing transitions (for both protons and neutrons), while the thick arrows describe the shape transitions.  The spherical nuclei $^{144-146}$Nd do not support rotational bands, so the enhancement factor $K$ is due to vibrational states alone. We observe that this vibrational enhancement factor decays to $\sim 1$ at excitation energies above the proton and neutron pairing transitions.  The heavier neodymium isotopes $^{148-152}$Nd are deformed in their ground state.  We observe that $K$ has a local minimum above the pairing transitions, which we attribute to the interplay between the decay of vibrational collectivity and the enhancement due to rotational collectivity. The plateau that follows the local minimum describes the enhancement from rotational states. This rotational enhancement decays in the vicinity of the shape phase transition.

\section{Conclusion}
The SMMC is a powerful method for calculating statistical and collective properties of nuclei in the framework of the CI shell model in very large model spaces. We reviewed a recent method to circumvent the odd-particle sign problem and its application to the calculation of pairing gaps and level densities of odd-mass nuclei. We discussed recent applications of SMMC to heavy rare-earth nuclei and in particular the microscopic emergence of collectivity in such nuclei. 

\section*{Acknowledgments}

It is a pleasure to dedicate this article to Aldo Covello in recognition of his important contributions to nuclear structure physics. This work was supported in part by the DOE grant No.~DE-FG-0291-ER-40608,  by the JSPS Grant-in-Aid for Scientific Research (C) No.~25400245, and by the Turkish Science and Research Council (T\"{U}B\.{I}TAK) grant No.~ARDEB-1001-112T973. Computational cycles were provided by the NERSC high performance computing facility and by the High Performance Computing Center at Yale University.

\end{document}